\documentstyle[aps,preprint]{revtex}
\begin{document}
\draft
\title{
Dynamical properties of the single--hole $t$--$J$ model\\
on a 32--site square lattice
}
\author{P. W. Leung$^*$}
\address{Physics Department, Hong Kong University of Science
and Technology,\\ Clear Water Bay, Hong Kong}
\author{R. J. Gooding$^\dagger$}
\address{Department of Physics and the Centre for Materials Science
and Engineering,\\
Massachusetts Institute of Technology, Cambridge, MA 02139}
\date{\today}
\maketitle
\begin{abstract}
We present results of an exact diagonalization calculation
of the spectral function
$A(\bf k, \omega)$ for a single hole described by the $t$--$J$
model propagating on
a 32--site square cluster. The minimum energy state is found at
a crystal momentum
${\bf k} = ({\pi\over 2}, {\pi\over 2})$, consistent with theory,
and our measured dispersion relation agrees well with that determined
using the
self--consistent Born approximation.  In contrast to smaller
cluster studies, our
spectra show no evidence of string resonances.
We also make a qualitative comparison of the variation of the spectral
weight in various regions of the first Brillouin zone with recent ARPES data.
\end{abstract}

\vskip 1.0 truecm

The $t$--$J$ model has received a lot of attention in recent years.
It is believed to be the simplest strong--coupling model of the low energy
physics of the anomalous metallic state of high--temperature
superconductors \cite {anderson,zhangrice}.  The Hamiltonian of the model is
\begin{equation}
{\cal H} = -t\sum_{\langle ij\rangle\sigma}(\tilde{c}^\dagger_{i\sigma}
\tilde{c}_{j\sigma}+{\rm H.c.})+J\sum_{\langle ij\rangle}
({\bf S}_i\cdot {\bf S}_j
-\frac{1}{4}n_in_j),
\label{hamiltonian}
\end{equation}
where $\langle ij\rangle$ denotes nearest neighbor sites,
and $\tilde{c}^\dagger_{i\sigma}$, $\tilde{c}_{i\sigma}$ are the constrained
operators, $\tilde{c}_{i\sigma}=c_{i\sigma}
(1-c^\dagger_{i,-\sigma}c_{i,-\sigma})$.

Although understanding the $t$--$J$ model doped with many
holes is an important
issue in the (potential) resolution of the high $T_c$ mystery,
the single--hole
state is by itself an interesting and important problem.
For example, it leads to the many--body wave functions which are
the starting point of any rigid--band filling analysis \cite {trugman}.
Further, the single--hole model has been studied in great detail by various
analytical and numerical theoretical techniques \cite {dagotto94},
and recently angle--resolved photoemission (ARPES)
data for the insulating, antiferromagnetically--ordered
CuO$_2$ planes in Sr$_2$CuO$_2$Cl$_2$ have become available \cite {wells95},
thus allowing for detailed comparisons between theory and experiment.
To be specific, these results on the  properties of a single hole
in the CuO$_2$ plane provide a direct test of how well the $t$--$J$ model
(or any other microscopic Hamiltonian)
describes the low energy physics of the CuO$_2$ plane \cite {nazarenko}.

In this paper we report the first exact diagonalization results, found using
the Lanczos algorithm, for a single hole described by the $t$--$J$ model on a
{\bf 32--site} square lattice. We use $t$ as the unit of energy, i.e., $t = 1$.
Figure \ref{qspace} shows the distinct {\bf k} points in the
reciprocal space of the 32--site square lattice.
Previous calculations for this model were mostly done on the
16--site ($4\times4$)
square lattice, where the {\bf k} points along the antiferromagnetic Brillouin
zone (ABZ) edge (from $(0,\pi)$ to $(\pi,0)$) are degenerate.
Other square lattices
that have been studied (18--, 20--, and 26--site)
do not have the important $\bf k$ points along the ABZ edge,
viz., the single--hole ground state wavevector $(\frac{\pi}{2},\frac{\pi}{2})$
nor many points along the $(1,1)$ direction (from $(0,0)$ to $(\pi,\pi)$).
The 32--site square lattice is the smallest one which has these
high symmetry points, and does not have the spurious degeneracy
of the $4\times4$ square lattice. Thus, this paper represents
a major advance in
the exact, unbiased, numerical treatment of an important
strong--coupling Hamiltonian.

In order for us to complete the exact diagonalization on such a large lattice,
we use translation and one reflection symmetry to reduce the total
number of basis states to about 150 million.
At ${\bf k}=(\frac{3\pi}{4},\frac{\pi}{4})$,
no reflection symmetry can be used and the total number of basis
states is about 300 million.
To study the effect of finite system sizes, we will supplement
our results with data obtained from smaller systems: the
$N=16$ $(4\times4)$ cluster,
as well as a 24--site $(\sqrt{18}\times\sqrt{32})$
cluster that includes many of
the important wave vectors \cite {24site}.

The electron spectral function is defined by
\begin{equation}
A({\bf k},\omega)=
\sum_n |\langle\psi^{N-1}_n|\tilde{c}_{{\bf k},\sigma}|\psi^N_0
\rangle|^2 \delta(\omega-E^N_0+E^{N-1}_n),
\end{equation}
where $E^N_0$ and $\psi^N_0$ are the ground state energy and wavefunction
of the model at half filling, respectively,
and $E^{N-1}_n$ and $\psi^{N-1}_n$ are
the energy and wavefunction of the $n$th eigenstate of the single--hole state,
respectively.  $A({\bf k},\omega)$ is calculated using a continued fraction
expansion \cite{llmno92} with 300 iterations and
an artificial broadening factor $\epsilon=0.05$.
We obtain $A({\bf k}, \omega)$ that are well converged using these
quantities.

Figure \ref{aq}(a) shows $A({\bf k},\omega)$ at $J=0.3$ from $(0,0)$
to $(\pi,\pi)$.  At $(0,0)$, the spectrum has a quasiparticle
peak at $\omega \sim 1.34$ and a broad feature at lower energies.
As ${\bf k}$ moves away from $(0,0)$ along the $(1,1)$ direction
towards $(\pi,\pi)$, spectral weight shifts from the broad feature
to both a higher energy quasiparticle peak and the low energy
tail of the spectrum.  The quasiparticle peak increases in intensity
and shifts to higher energies, reaching the valence band maximum at
$(\frac{\pi}{2},\frac{\pi}{2})$.  When {\bf k} goes
further towards $(\pi,\pi)$, the quasiparticle peak moves to lower
energies and its intensity drops significantly.
Spectral weight move towards the
central part of the spectrum again, eventually leaving only a very small
quasiparticle peak at $\omega=1.2313$, and a broad low energy structure.
Figure \ref{aq}(b) shows $A({\bf k},\omega)$ at other distinct {\bf k}.

 From Fig.~\ref{aq} one sees that along the ABZ edge $A({\bf k},\omega)$
are qualitatively similar \cite{mrrv91}.  They have strong quasiparticle peaks
which do not disperse much.  The intensity of the quasiparticle peak
is the largest at $(\pi,0)$.  As {\bf k} moves from $(\pi,0)$ to
$(\frac{\pi}{2},\frac{\pi}{2})$, the intensity of
the quasiparticle peak decreases
and is the smallest at $(\frac{\pi}{2},\frac{\pi}{2})$ along the
ABZ edge.
This can be made quantitative by calculating
the quasiparticle weight, which is defined by
\begin{equation}
Z_{\bf k}=\frac{|\langle\psi^{N-1}_m|\tilde{c}_{{\bf k}\sigma}|
\psi^N_0\rangle|^2}
{\langle\psi^N_0|\tilde{c}^\dagger_{{\bf k}\sigma}
\tilde{c}_{{\bf k}\sigma}|\psi^N_0\rangle},
\end{equation}
and is proportional to the area under the quasiparticle peak.
Table~\ref{zq} shows the values of $Z_{\bf k}$.
$Z_{\bf k}$ is the largest at $(\pi,0)$,
and remains large along the edge of the ABZ.
Outside the ABZ, $Z_{\bf k}$ decreases very fast, especially
along the $(1,1)$ direction.  $(\frac{\pi}{2},\frac{\pi}{2})$
is a saddle point:
$Z_{\bf k}$ is a maximum
along the $(1,1)$ direction, but a minimum along the direction
from $(0,\pi)$ to $(\pi,0)$.

 From our spectral function results we have extracted a
quasiparticle dispersion relation,
and in Fig.~\ref{dispersion} we display the resulting $E({\bf k})$,
which is the location of the quasiparticle
peak of $A({\bf k},\omega)$.  The numerical values of $E({\bf k})$
are tabulated in Table~\ref{zq}.  In agreement with earlier
(and smaller system size)
results, we find that the band maximum and minimum are at
$(\frac{\pi}{2},\frac{\pi}{2})$ and $(\pi,\pi)$,
respectively \cite {disp}.
Further, the valence band maximum is located at ${\bf k} = ({\pi\over 2},
{\pi\over 2})$, consistent with the large theoretical effort that was applied
to the single--hole problem \cite {dagotto94}.
The solid line in Fig.~\ref{dispersion} is the
self--consistent Born approximation (SCBA)
result on a $16\times16$ square lattice \cite{mh91}.
The agreement between our 32--site dispersion relation
and that of the SCBA calculation
is encouraging.

To study the finite--size effects in our results, we can compare our cluster
to smaller system size results.  Unfortunately, a
comparison of the spectral function
is limited by the availability of the particular
{\bf k} points in smaller lattices.
When comparing $A({\bf k},\omega)$ of the 16-- \cite{dagotto90,poilblanc93},
18--, 20--, 26-- \cite{poilblanc93}, and 32--site lattices
at $(0,0)$ and $(\pi,\pi)$ \cite{convention},
we find that the high energy features (including the quasiparticle peak)
are not very sensitive to the system size. However, the low energy
features are smeared out and broadened in larger systems ---
this point was also made in Ref.~\onlinecite{poilblanc93}.
We also calculated the spectral function for the
24--site lattice \cite {24site}.
Although it is not square, this cluster has the same five {\bf k}
points along the $(1,1)$ direction as the 32--site lattice.
$A({\bf k},\omega)$ at these points are qualitatively similar to those
of the 32--site results, except for detailed values of the intensity.
We conclude that the 24--site lattice is large enough to capture
the essential shape of the spectral function along the $(1,1)$
direction, while the 16--site lattice is too small, especially
when the lower energy features are concerned.
In particular, the well--defined secondary peaks
found at $(\frac{\pi}{2},\frac{\pi}{2})$
on the 16--site lattice, which were interpreted to be related
to the ``string picture'' \cite{dagotto90}, are not found in our
32--site system.  Hence we find no evidence supporting the string
picture (at least for this value of $J$).
Further, the single--hole energy, defined as
$E_h=E^{N-1}_0-E^N_0=-E(\frac{\pi}{2},\frac{\pi}{2})$,
is calculated at $0.1\le J\le0.8$ for the 32--site system.
Fitting to the form $E_h-J=a+bJ^\nu$
gives $a=-3.24$, $b=2.65$, and $\nu=0.72$.
This is consistent with the 16--site results \cite{dagotto90},
$a=-3.17$, $b=2.83$, and $\nu=0.73$, and also with the large
cluster estimate of the SCBA calculations \cite{mh91}.
However, $\nu$ is not $\frac{2}{3}$ as suggested
in the string picture \cite{lm92}.

Figures~\ref{bw} and \ref{zqN}
show the bandwidth $W=E(\frac{\pi}{2},\frac{\pi}{2})-E(\pi,\pi)$
and the quasiparticle
weight $Z_{(\frac{\pi}{2},\frac{\pi}{2})}$
at different $J$ for $N=16$, 24 and 32.
Fitting to the functional forms $a+bJ^\nu$ is more difficult.
The best estimates of the coefficients are shown in the graphs.
(The negative y--intercept of $W$ versus $J$ is non--physical and has
been accounted for by higher order processes in $t$ \cite{shsz90}.)
For $J\le0.4$, $W$ can be fitted reasonably well by a linear relation
in $J$.
Also, we find that $Z_{(\frac{\pi}{2},\frac{\pi}{2})}$ does not scale
monotonically with $N$ in this range of $J$.  Consequently, we feel that a
precise extrapolation to large $N$ based on available
numerics is not justified.
However, our results support the hypothesis that at the $J$ value of physical
interest ($J \sim 0.3$) $Z_{(\frac{\pi}{2},\frac{\pi}{2})}$
remains non--zero for macroscopic $N$.

Finally, our results also allow for a comparison to the
experimental dispersion relation data
of Wells {\em et al.} \cite {wells95}. It is clear that the
flat dispersion relation that we find
along the ABZ does not agree with the measured $E({\bf k})$,
in direct contradiction
to a recent theory of the ARPES spectra based on the
$t$--$J$ model \cite {laughlin}.   The behavior of
$Z_{\bf k}$ along the ABZ edge also differs
from experimental results: $Z_{\bf k}$ is a minimum at
${\bf k}=(\frac{\pi}{2},\frac{\pi}{2})$
along this direction, while the intensity of the ARPES peak is a
maximum.
Our exact,
unbiased numerical results clearly demonstrate that the
$t$--$J$ model cannot explain
the data. However, as was shown in previous work
\cite {nazarenko,24site}, it is now known
that a $t^\prime$ is required to describe hole motion via a single--band model
in a CuO$_2$ plane.
Thus, we defer a more quantitative comparison
of theory and the dispersion relation obtained by
experiment until a future publication
wherein the effect of  $t^\prime$ will be presented.
A qualitative comparison to the
variation of spectral weight as a function of ${\bf k}$ found numerically
to that found in the ARPES data shows one surprising agreement.
To be specific, note that
the $t$--$J$ model is a strong--coupling model which when
undoped displays long--ranged
antiferromagnetic broken symmetry. Thus one expects that
for only one hole, there should
be an equivalence of $A({\bf k},\omega)$ under a shift of
a reciprocal lattice vector
of the magnetic lattice. As seen in Fig.~\ref{aq} our results
clearly do not display this feature.
However, neither do the ARPES experiments \cite {wells95},
and thus, unlike the original
conjecture of Ref. \onlinecite {wells95}, the strength of
the on--site correlations cannot
necessarily be resolved via photoemission.

Summarizing, we have presented the spectral function of
a single hole propagating
on a 32--site square lattice described by the $t$--$J$ model. These exact,
unbiased, numerical data now serve as an acid test of analytical theories of
this important strong--coupling Hamiltonian.

\acknowledgments

We wish to thank T. K. Ng, Ken Vos, and Barry Wells for helpful comments.
This work was supported by Hong Kong RGC grant HKUST619/95P (PWL)
and the NSERC of Canada (RJG).
Numerical diagonalizations of the 32--site system were performed on the
Intel Paragon at HKUST.

\vskip 2.0 truecm
$^*$ Email: phleung@usthk.ust.hk

$^\dagger$ Permanent Address: Dept. of Physics, Queen's University,
Kingston, ON Canada K7L 3N6.

% now the references. delete or change fake bibitem. delete next three
%   lines and directly read in your .bbl file if you use bibtex.

\begin{figure}
\caption{(a) The 32--site square lattice employed in this study.
(b) The distinct {\bf k} points in the reciprocal space of this lattice.}
\label{qspace}
\end{figure}

\begin{figure}
\caption{The electron spectral function $A({\bf k},\omega)$ for $J=0.3$.}
\label{aq}
\end{figure}

\begin{figure}
\caption{
The quasiparticle dispersion relation of the 32--site
$t$--$J$ model at $J=0.3$.
The solid line is from the SCBA calculation of Ref.~\protect\onlinecite{mh91},
but with a vertical offset to make $E(\frac{\pi}{2},\frac{\pi}{2})$
agree with ours.}
\label{dispersion}
\end{figure}

\begin{figure}
\caption{The bandwidth of the $t$--$J$ model at various $J$ for
$N=16$, 24, and 32.  The lines represent the best fits with the
indicated functional form.}
\label{bw}
\end{figure}

\begin{figure}
\caption{
The same as in Fig.~\protect\ref{bw} but for the quasiparticle weight
$Z_{(\frac{\pi}{2},\frac{\pi}{2})}$.
}
\label{zqN}
\end{figure}

\begin{table}
\caption{Quasiparticle weight $Z_{\bf k}$ and energy $E({\bf k})$
of the 32--site
$t$--$J$ model at $J=0.3$.
The half--fill ground state energy
$E^N_0$ is $-11.329720$. }
\label{zq}
\begin{tabular}{r@{}lcc}
\multicolumn{2}{c}{\bf k}&$Z_{\bf k}$&$E({\bf k})$\\
\tableline
(0,&0)                                  & 0.145200 & 1.339454 \\
($\frac{\pi}{4}$,&$\frac{\pi}{4}$)      & 0.234239  & 1.563812 \\
($\frac{\pi}{2}$,&$\frac{\pi}{2}$)      & 0.311065 & 1.832213 \\
($\frac{3\pi}{4}$,&$\frac{3\pi}{4}$)    & 0.098759 & 1.509999 \\
($\pi$,&$\pi$)                          & 0.005499  & 1.231307 \\
($\pi$,&$\frac{\pi}{2}$)                & 0.211046  & 1.672890 \\
($\pi$,&0)                              & 0.341747 & 1.710318 \\
($\frac{\pi}{2}$,&0)                    & 0.295290  & 1.619414 \\
($\frac{3\pi}{4}$,&$\frac{\pi}{4}$)     & 0.320631  & 1.777018 \\
\end{tabular}
\end{table}


\begin{references}
\bibitem{disp} Note that at $(0,0)$ and $(\pi,\pi)$,
$E({\bf k})$ are not the lowest energy states in
their respective symmetry sectors.
The lowest energy states with these momenta are not $S=\frac{1}{2}$
states, and hence do not appear in the spectral function.
Also, although the quasiparticle peak that we find at
$(\pi,\pi)$ is extremely small,
$E(\pi,\pi)$ is given very accurately by the Lanczos iterations.
\bibitem{convention} Note that we report the electron spectrum
which is related to the hole spectrum in Refs.~\onlinecite{dagotto90}
and \onlinecite{poilblanc93} by the transformations
${\bf k}\rightarrow{\bf k}+(\pi,\pi)$ and $\omega\rightarrow -\omega$.
\end{references}
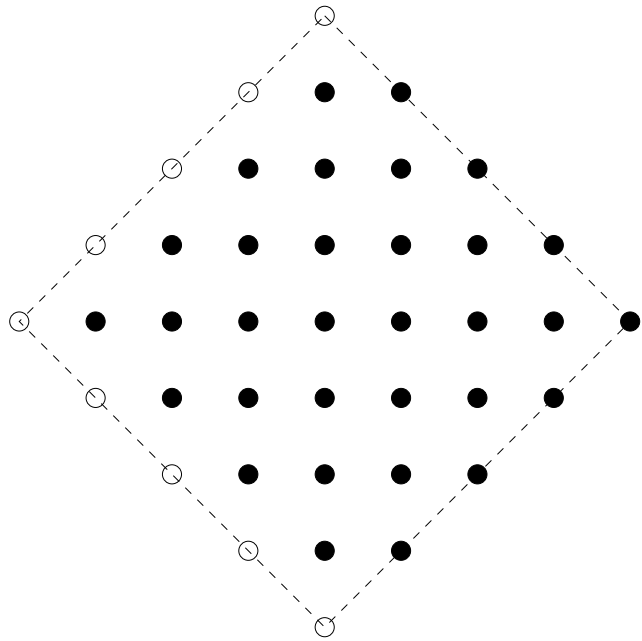
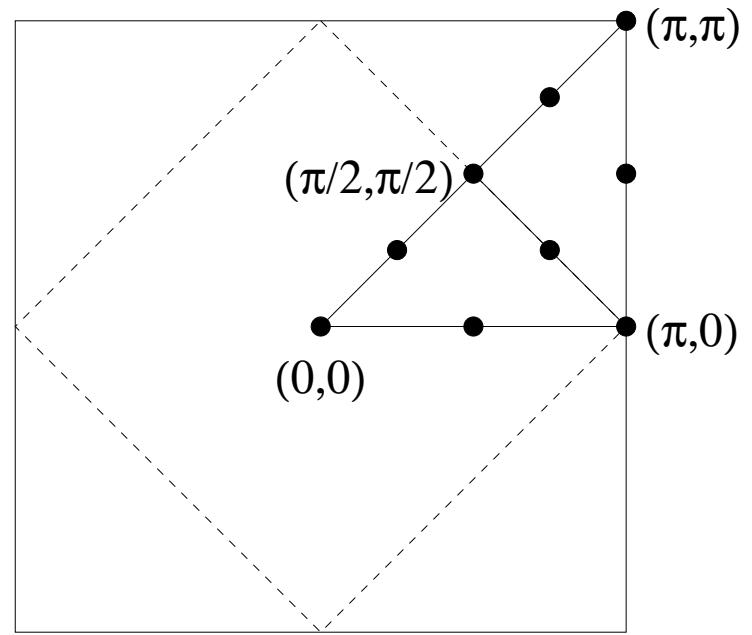
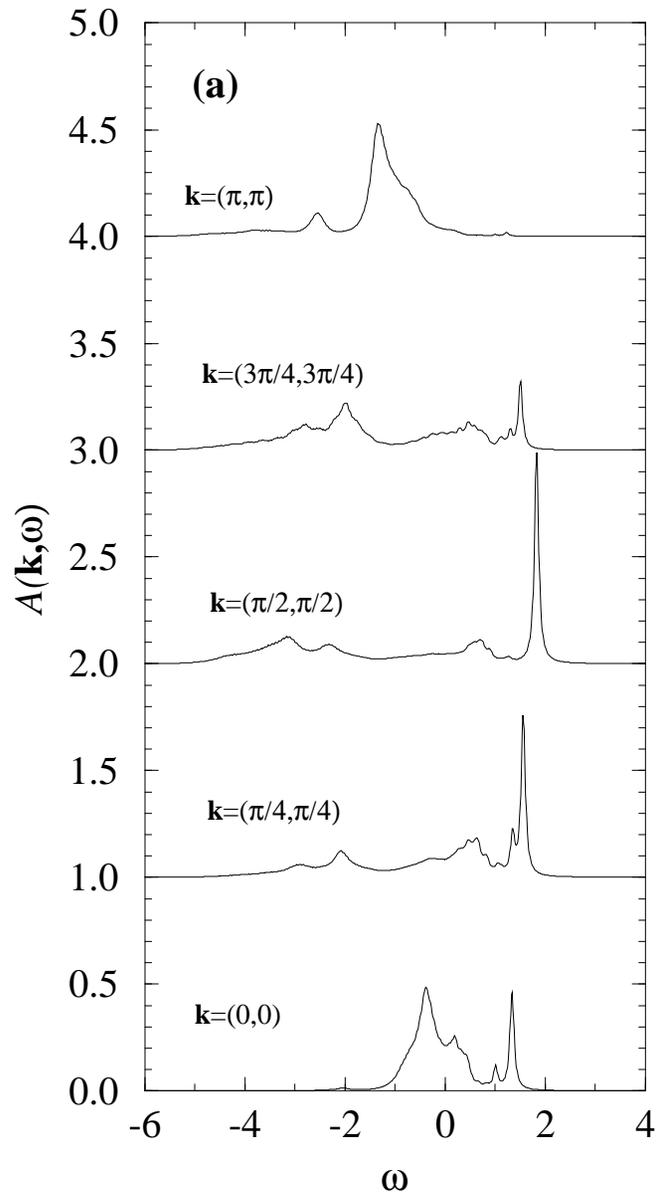
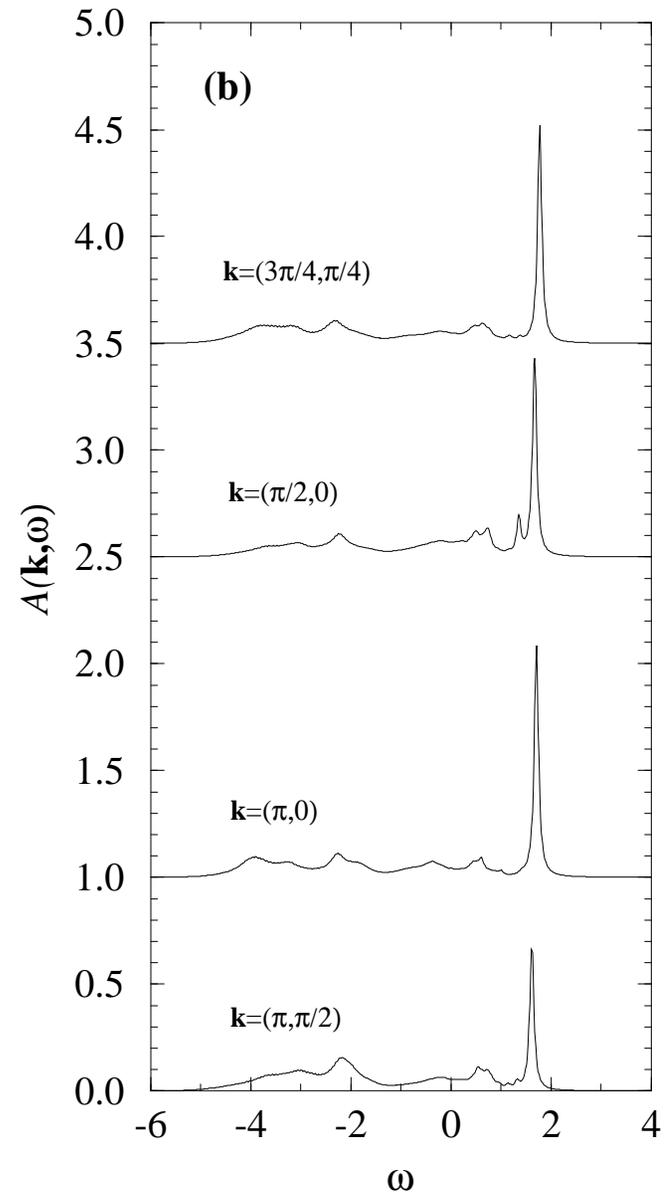
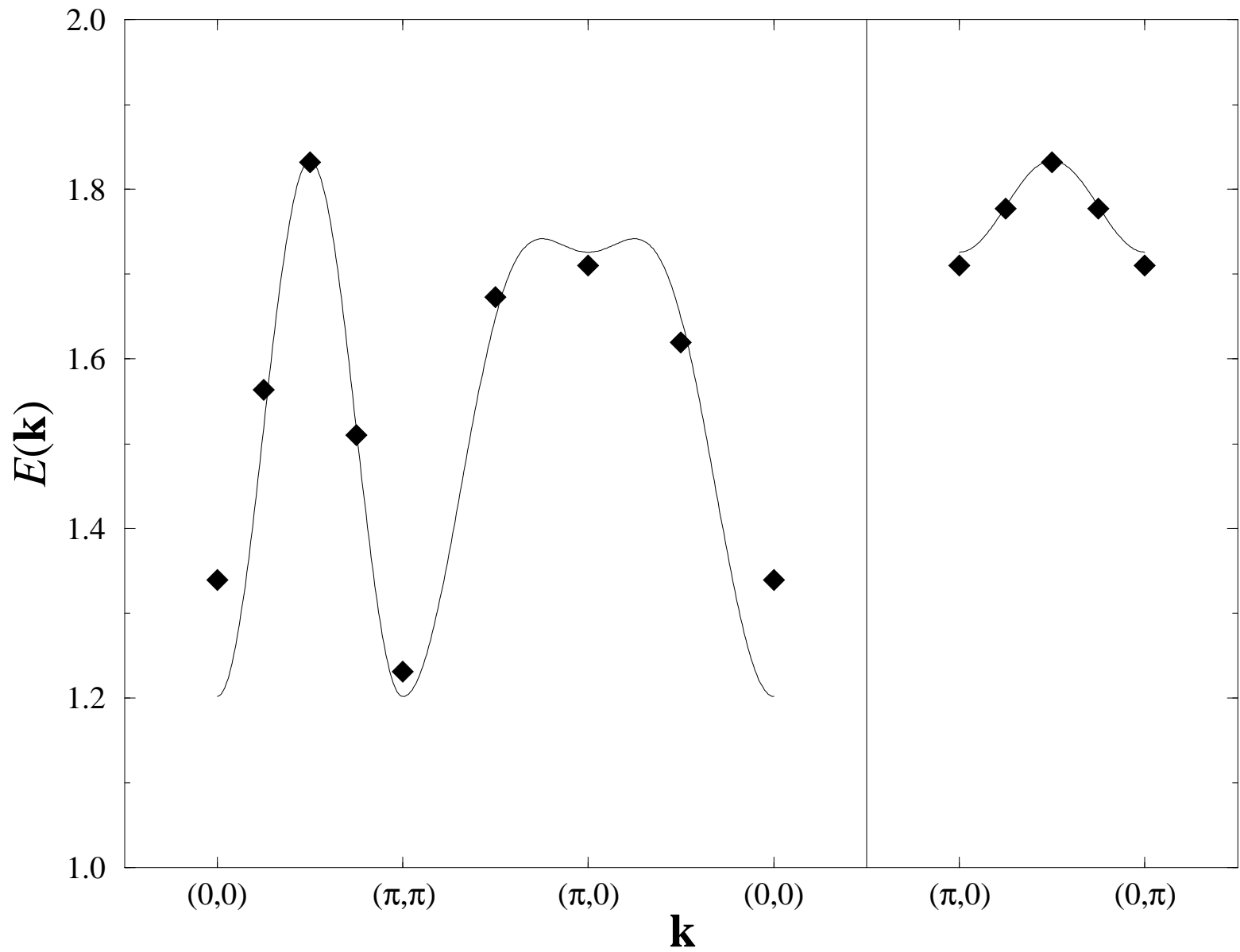
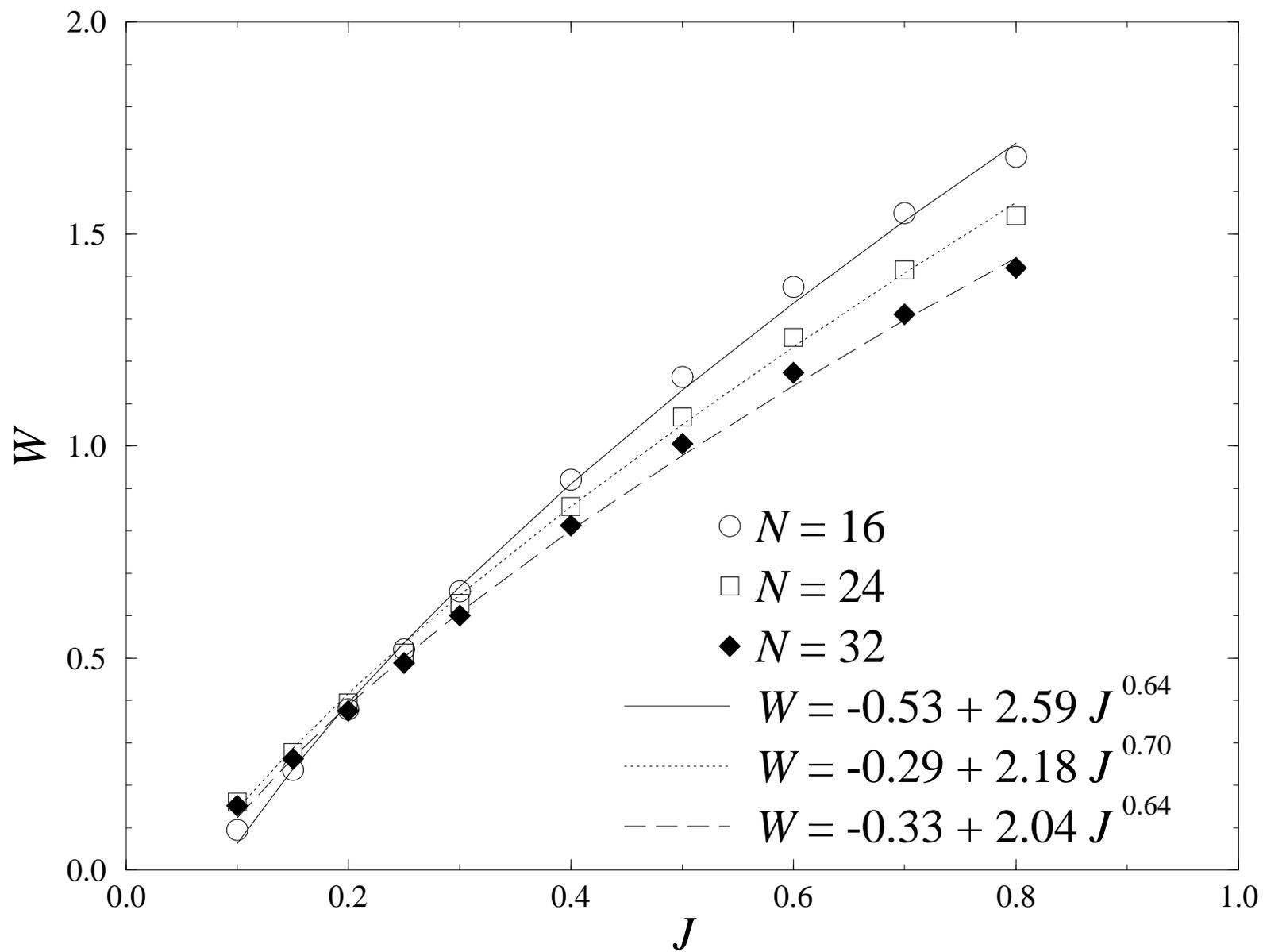
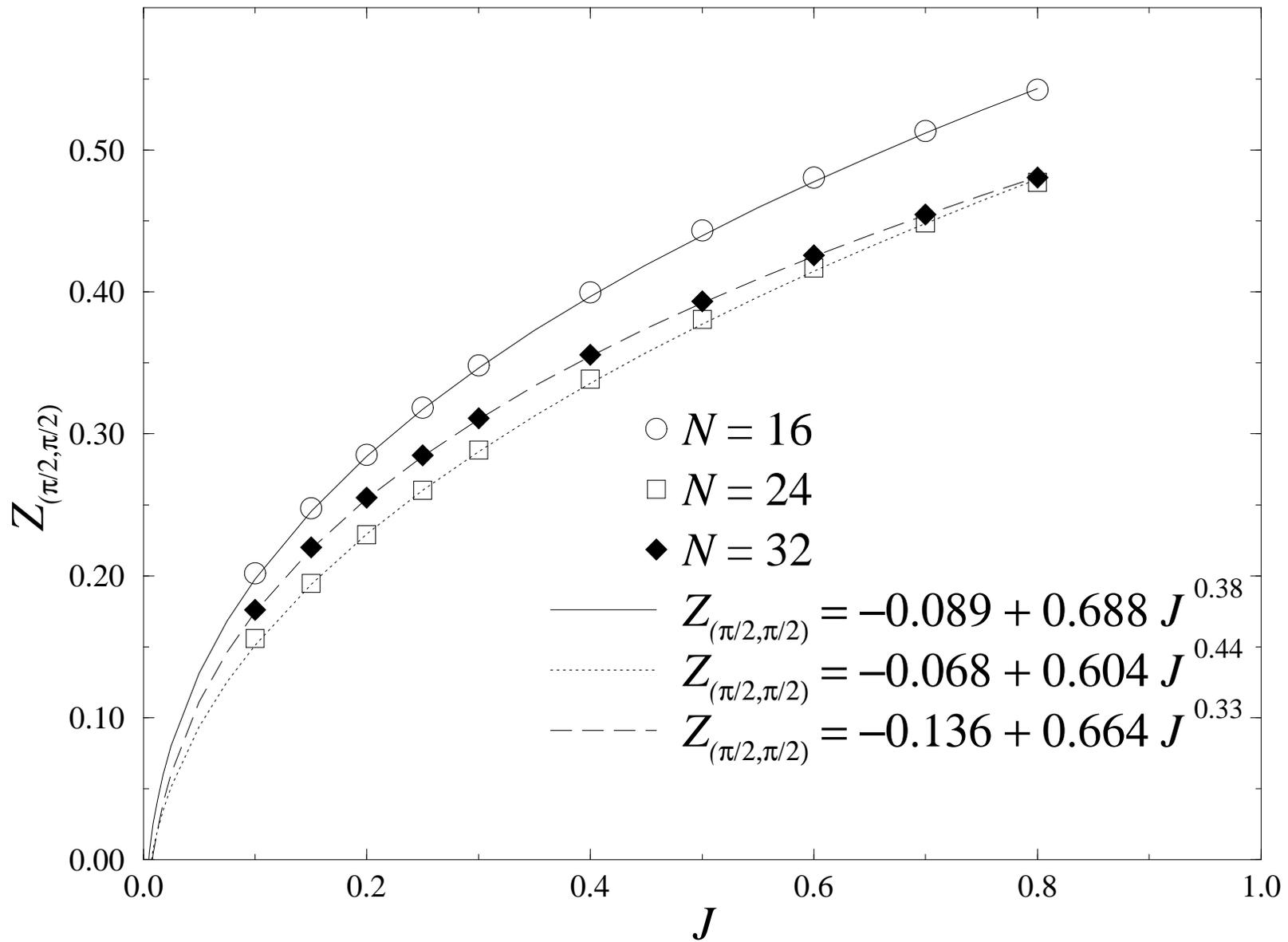
\end{document}